\documentclass[mypaper,7pt,twoside]{CoAst}
\usepackage{epsf,graphicx,fancyhdr}
\renewcommand{\chaptermark}[1]{\markboth{#1}{}}

\newcommand{\mnras}{MNRAS}

\newcommand{\kopf}{\small\itshape Comm. in Asteroseismology \\ Contribution to the Proceedings of the 38$^{th}$\,LIAC\,/\,HELAS-ESTA\,/\,BAG, 2008
}

\newcommand{\Authors}[1]{\begin{center}\normalsize\bf\sf #1 \end{center}}

\renewcommand{\author}[1]{\begin{center}\normalsize\bf\sf #1 \end{center}}
\newcommand{\Address}[1]{\begin{center}\small\sf #1 \end{center}}

\DeclareGraphicsExtensions{.eps,.jpg}

\newcommand{\Session}[1]{{\vspace{3mm}\small \noindent  \hspace*{3mm} Session: } #1 \normalsize}

	\newcommand{\four}{\small Observed frequencies in pulsating massive stars}

\renewenvironment{abstract}{\section*{Abstract}\normalsize\sf}{}
\newcommand{\References}[1]{\begin{flushleft}{\large References\\}\vspace*{2mm}\small #1 \end{flushleft}}

\newcommand{\chapterCoAst}[2]{\chapter[\sf\normalsize #1\\ \footnotesize \hspace*{5mm}by #2 \sf\normalsize][]{#1\\}\rhead[\fancyplain{}{\sf\footnotesize \center{#1}}]{\fancyplain{}{\sffamily\thepage}}\lhead[\fancyplain{\kopf}{\sffamily\thepage}]{\fancyplain{\kopf}{\sf\footnotesize \center{#2}}}}

\newcommand{\figureCoAst}[5]{\begin{figure}[#4]
\centering
\includegraphics*[#5]{#1}
\caption{#2}
\label{#3}
\end{figure}}

\renewcommand{\chaptername}{}

\def\rfr{\smallskip\par\noindent
        \hangindent=7truemm
        \hangafter=1}

\begin{document}
\sf

\chapterCoAst{More on pulsating B-type stars in the Magellanic Clouds}
{P.D.~Diago, J.~Guti\'{e}rrez-Soto, J.~Fabregat and C.~Martayan} 
\Authors{P.D.~Diago$^{1}$, J.~Guti\'{e}rrez-Soto$^{1,2,3}$, J.~Fabregat$^{1,2}$ and C.~Martayan$^{2,4}$}
\Address{
$^1$ Observatori Astron\`{o}mic de la Universitat de Val\`{e}ncia, Ed. Instituts d'Investigaci\'{o}, \\Pol\'{i}gon La Coma, 46980 Paterna, Val\`{e}ncia, Spain\\
$^2$ GEPI, Observatoire de Paris, CNRS, Universit\'e Paris Diderot, \\Place Jules Janssen 92195 Meudon Cedex, France\\
$^3$ LESIA, Observatoire de Paris, CNRS Universit\'e Paris Diderot, \\ Place Jules Janssen 92195 Meudon Cedex, France\\
$^4$ Royal Observatory of Belgium, 3 Avenue Circulaire, B-1180 Brussels, Belgium
}

\noindent
\begin{abstract}


We present here the results of our research for B-type pulsators in low metallicity environments, searching for short-term periodic variability in a large sample of B and Be stars in the Magellanic Clouds (MC), for which the fundamental astrophysical parameters were accurately determined. A significant number of $\beta$ Cephei and SPB-like pulsators at low-metallicity have been detected, conflicting with the current theoretical models of pulsation. In addition, we have placed these pulsating stars in the HR diagram mapping the observational instability regions for the MC metallicities. The large sample of B and Be stars analysed allows us to make a reliable statistics of the pulsating B-type stars in the MC. Finally, we have made a comparison between pulsational theory and observations in low metallicity environments.

\end{abstract}

\Session{ \four } 


\section*{Introduction}

A significant fraction of main-sequence B-type stars are variable. The whole main-sequence in the B spectral domain is populated by two classes of pulsators: the $\beta$ Cephei stars
and the Slowly Pulsating Stars (SPB). 
Pulsations in $\beta$ Cephei and SPB stars are due to the $\kappa$-mechanism activated by the metal opacity bump
. $\beta$ Cephei stars do pulsate in low-order p- and g-modes with periods similar to the fundamental radial mode. SPB stars are high-radial order g-mode pulsators with periods longer than the fundamental radial one. \textit{Pamyatnykh (1999)} showed that the $\beta$ Cephei and SPB instability strips practically vanish at $Z < 0.01$ and $Z < 0.006$, respectively. The metallicity of the Magellanic Clouds (MC) has been measured to be around $Z= 0.002$ for the Small Magellanic Cloud (SMC) and $Z= 0.007$ for the Large Magellanic Cloud (LMC) (see \textit{Maeder et al. 1999} and references therein). Therefore, it is  expected to find a very low occurrence of $\beta$ Cephei and SPB pulsators in the LMC and no pulsating B-type stars in the SMC. 
Recently, new B-type pulsators have been found in low-metallicity environments (e.g. 
\textit{Karoff et al. 2008} or \textit{Diago et al. 2008}), suggesting that pulsations are still driven by the $\kappa$-mechanism even in low metallicity environments.

The new models provided by \textit{Miglio et al. (2007a,b)} based on OPAL and updated OP opacities have shown that the blue border of the SPB instability strip is displaced at higher effective temperatures at solar metallicity, and that a SPB instability strip exists at metallicities as low as $Z=0.005$. Their calculations however, do not predict $\beta$ Cephei pulsations at $Z=0.005$.

Another large class of stars populating the B-type main-sequence are the Be stars
. They are defined as non-supergiant B stars whose spectrum has displayed at least once emission lines mainly in the Balmer series. Emission lines comes from a circumstellar disk created by episodic matter ejections from the central star. In the Milky Way (MW), they show short-term variations like $\beta$ Cephei or SPB stars. Be stars are also known to be fast rotating stars.

Here we present the analysis of a sample of 128 B and Be stars from the LMC and 313 B and Be stars from SMC, for which \textit{Martayan et al. (2006, 2007)} provided accurate fundamental astrophysical parameters.

\section*{Results}

The search for periodic variability has been done by analysing the photometric time series provided by the MACHO project. The frequency analysis was done with the self-developed code \texttt{pasper}. The criterion used to determine whether the frequencies are statistically significant is the \textit{signal to noise amplitude ratio requirement} described in \textit{Breger et al. (1993)}. Moreover, once we had the significant frequencies, we performed a visual inspection of the phase diagrams for each star folded with the detected frequencies. All the details concerning data and methods used can be found in \textit{Diago et al. (2008)}.

Many of the short-period variables have been found multi-periodic and some of them show the beating phenomenon due to the presence of close frequencies. This effect is a signature of non-radial pulsations.

\subsection*{Small Magellanic Cloud}

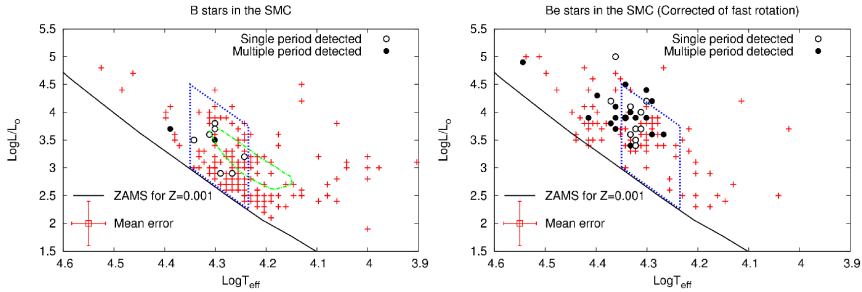
\figureCoAst{SMC-AstroPH}{Location of the B (left) and Be (right) samples of the SMC in the the HR diagram: single crosses represent stars in our sample, the empty circles represent single period detection and the filled ones multiple period detection. The dashed line delimits the suggested SPB instability strip for the SMC. In the left panel we have depicted with dash-point-dashed line the SPB instability strip computed by \textit{Miglio et al. (2007b)} at $Z=0.005$ with OP opacities. 
}{SMC}{h}{clip,angle=0,width=115mm}

In the left panel of Fig.~\ref{SMC} we show the position in the HR diagram for the 9 short-period variable B stars found. All pulsating B stars are restricted to a narrow range of temperatures. Moreover, all stars but one have periods longer than 0.5 days, characteristic of SPB stars. Thus, we suggest an observational SPB instability strip for the SMC metallicity, that it is shifted towards higher temperatures than for the MW. We propose the hottest pulsating star in our sample to be a $\beta$ Cephei variable,  because it has two close periods in the range of p-mode Galactic pulsators. If it is indeed a $\beta$ Cephei star, therefore it would constitute an unexpected result, as the current stellar models do not predict p-mode pulsations at the SMC metallicities (see \textit{Miglio et al. 2007a,b}).

A puzzling circumstance regarding our sample of SPB stars is that only one has been detected as multi-periodic. So, the variability in some of these stars could not be caused by pulsations, but by other phenomena like eclipsing binaries or ellipsoidal binarity. However, none of these stars have been detected as binary in the spectroscopic survey of \textit{Martayan et al. (2007)}. Thus, we consider our figure of eight stars as an upper limit to the number of bona-fide SPB stars in our sample.

We have represented in the right panel of Fig.~{\ref{SMC}} the 32 pulsating Be stars in the HR diagram (using $\Omega/\Omega_{c}=95\%$). In the case of Be stars it is more difficult to distinguish the pulsational type by using the detected frequencies, because the fast rotation effects affect the observed periods
. We have included the suggested SPB region as described above in the figure and as most of the Be stars are located inside or very close to this region, suggesting that they are g-mode SPB-like pulsators. Three stars are significantly outside the strip towards higher temperatures, all of them multi-periodic, with periods lower than 0.3 days. Therefore, we propose that these stars are probably $\beta$ Cephei-like pulsators.

The detected frequencies with their amplitude and phases, the phase diagrams, beating of close frequencies and the detailed discussion for the SMC results are published in \textit{Diago et al. (2008)}.

\subsection*{Large Magellanic Cloud}

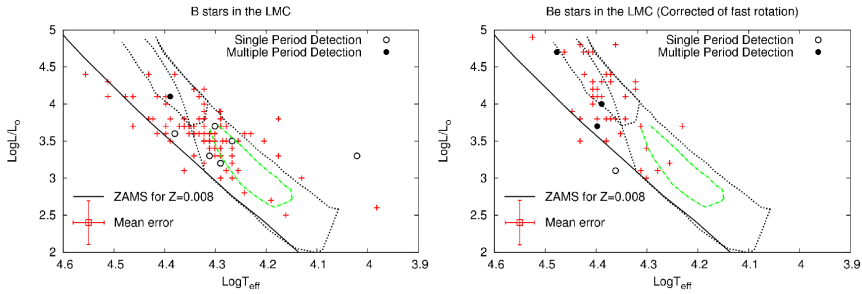
\figureCoAst{LMC-AstroPH}{Location of the B (left) and Be (right) samples of the LMC in the the HR diagram: single crosses represent stars in our sample, the empty circles represent single period detection and the filled ones multiple period detection. The dashed line delimits the $\beta$ Cephei and SPB instability strips at $Z=0.01$ and the dash-point-dashed line the SPB instability strip at $Z=0.005$ computed by \textit{Miglio et al. (2007b)} with OP opacities. 
}{LMC}{h}{clip,angle=0,width=115mm}

In the LMC we have found 7 short-period variables among the B star sample (one multi-periodic). Their positions are displayed in the HR diagram in the left panel of Fig.~\ref{LMC}. The periods obtained for these stars are typically of Galactic $\beta$ Cephei stars except one. Concerning Be stars, we found 4 short-period variables (3 multi-periodic), depicted in the right panel of Fig.~\ref{LMC} (using $\Omega / \Omega_{c} = 85\%$). The periods obtained for these pulsating Be stars are compatible with Galactic $\beta$ Cephei, but as in the SMC, the high rotational rates prevents us from distinguish the pulsational type using the observed periods. It is remarkable that, as in the SMC, the hottest stars are those that are multi-periodic. Our work is ongoing and it will be published by \textit{Diago et al. (in preparation)}.

\section*{Discussion}


As mentioned in the introduction, one expects a lower occurrence of B-type pulsators in the low metallicity environments of the MC. On the other hand, according to \textit{Maeder \& Meynet (2001)}, the high rotational velocities favours the metal enrichment of the surface of fast rotating stars due to the rotational mixing. In this way, this metal enrichment could feed the pulsational mechanism on the fast rotating stars
. Moreover, \textit{Zorec et al. (2005)} and \textit{Fr\'{e}mat et al. (2005)} have observed on Be stars that these fast rotation effects appear in stars with a ratio of the angular rotational velocity with respect to the critical angular rotational velocity ($\Omega / \Omega_{c}$) much larger than $60\%$.

Concerning our sample of B stars in the MC, the ratios of angular velocities are measured to be about $\Omega / \Omega_{c} = 37\%$ for the LMC and $\Omega / \Omega_{c} = 58\%$ for the SMC (see \textit{Martayan et al. 2007}). These values are very close to the one for the Galactic B-type stars, which is about $\Omega / \Omega_{c} \sim 40\%$. In both cases, the ratios are lower than $60\%$. Therefore, we conclude that the B star samples of the LMC and SMC are not affected of fast rotation effects which can excite the pulsational mechanism. The B star samples are, consequently, only affected by the decreasing trend in metallicity that makes a lower fraction of observed B-type pulsators as we can see in our results (3rd row of Table~\ref{tabla2}).

In the case of Be stars, the decreasing trend in metallicity is also present (see last row of Table~\ref{tabla2}). However, Be stars are fast rotators and in addition they can rotate faster in lower metallicity environments, since the radiative winds are less efficient. The values of rotational velocity rates are about $\Omega / \Omega_{c} = 85\%$ for the LMC and $\Omega / \Omega_{c} = 95\%$ for the SMC (see \textit{Martayan et al. 2007}). They are very close to the one for the Galactic Be stars, which ranges from $83\%$ to $88\%$, depending on the authors. Consequently, the percentages of pulsating Be stars in our sample suggests that the fast rotation enhances the non-radial pulsations or amplifies the existing modes in the pulsating Be stars between MW/LMC and SMC. A similar result was obtained by \textit{Guti\'{e}rrez-Soto et al. (2007)} for Be stars in the MW.

\begin{table}
\caption{Percentages of short-period variables in the MC and in the MW compared with their rotational velocity rates and the metallicity of the stellar environment.}
\label{tabla2}
\centering
\begin{tabular}{l c c c}
\hline\hline
				& MW	& LMC		& SMC		\\
\hline
Metallicity			& 0.020	& 0.007		& 0.002		\\
\hline
$\Omega / \Omega_{c}$ (B stars)	& 40\%	& 37\%		& 58\%		\\
Pulsating B stars		& 16\%	& 6.9\%		& 4.9\%		\\
\hline
$\Omega / \Omega_{c}$ (Be stars)& 88\%	& 85\%		& 95\%		\\
Pulsating Be stars		& 74\%	& 15\%		& 25\%	\\
\hline
\end{tabular}
\end{table}

\section*{Conclusions}

The most important result in our investigation is the detection of $\beta$ Cephei and SPB-type pulsators in low metallicity environments, in contrast with the predictions of the current theoretical models. Pulsations in B stars seem to be damped by the decreasing trend of metallicity, so the lower the metallicity the lower the pulsations observed in these stars. For Be stars, the rapid rotation seems to enhance the presence of the non-radial pulsations or to amplify the existing modes, making them easier to be detected. As an alternative explanation, the prevalence of non-radial pulsations could be related to the yet unknown nature of the Be phenomenon, being in fact Z-enriched stars due to rotational mixing.


\section*{Questions}

\textit{A. Baglin:} Can you give statistical significances of the percentages you present? Are they $10\%$ or $50\%$?\\
\textit{P. D. Diago:} The samples analysed consist of 313 stars in the SMC and 128 stars in the LMC, so the size of both samples is large enough to consider the differences in Table~\ref{tabla2} to be significant, although we have not performed formal calculations of the statistical significance.\\
\\



\References{

\rfr Breger, M., Stich, J., Garrido, R. et al. 1993, A\&A, 271, 482
\rfr Diago, P. D., Guti\'{e}rrez-Soto, J., Fabregat, J. \& Martayan, C. 2008, A\&A, 480, 179
\rfr Fr{\'e}mat, Y., Zorec, J., Hubert, A.-M., \& Floquet, M. 2005, A\&A, 440, 305
\rfr Guti\'{e}rrez-Soto, J. , Fabregat, J., Suso, J., et al. 2007, A\&A, 476, 927
\rfr Karoff, C., Arentoft, T., Glowienka, L., Coutures, C., Nielsen, T.B., Dogan, G., Grundahl, F., \& Kjeldsen, H. 2008, \mnras, 386, 1085
\rfr Maeder, A.,  Grebel, E. K., \& Mermilliod, J.-C. 1999, A\&A, 346, 459
\rfr Maeder, A., \& Meynet, G.  2001, A\&A, 373, 555
\rfr Martayan, C., Floquet, M., Hubert, A.-M., et  al. 2006, A\&A, 452, 273
\rfr Martayan, C., Floquet, M., Hubert, A.-M., et  al. 2007, A\&A, 462, 683
\rfr Miglio, A., Montalb\'{a}n, J., \& Dupret, M.-A., et al. 2007, MNRAS, 375, L21
\rfr Miglio, A., Montalb{\'a}n, J., \& Dupret, M.-A. 2007, Communications in Asteroseismology, 151, 48
\rfr Pamyatnykh, A. A. 1999, Acta Astron., 49, 119
\rfr Zorec, J., Fr{\'e}mat, Y., \& Cidale, L. 2005, A\&A, 441, 235

}

\end{document}